\documentclass[a4paper,twoside,twocolumn,english,amssymb,aps,prb,reprint,twocolumn,superscriptaddress,floats,showkeys,showpacs]{revtex4-1}
\usepackage{graphicx,amsmath}
\usepackage{color}
\usepackage[utf8]{inputenc}

\usepackage{url}

\begin{document}

\title{Peculiar spectra of dark and bright excitons in alloyed nanowire quantum dots}
\author{M. Zieliński}
\email{mzielin@fizyka.umk.pl}
\affiliation{Institute of Physics, Faculty of Physics, Astronomy and Informatics, Nicolaus Copernicus University, Grudziadzka 5, 87-100 Torun, Poland}

\begin{abstract}
Excitons in alloyed nanowire quantum dots have unique spectra as shown here using atomistic calculations. The bright exciton splitting is triggered solely by alloying and despite cylindrical quantum dot shape reaches over $15~\mu$eV, contrary to previous theoretical predictions, however, in line with experimental data. 
This splitting can however be tuned by electric field to go below $1$~$\mu$eV threshold.
The dark exciton optical activity is also strongly affected by alloying reaching notable $1/3500$ fraction of the bright exciton and having large out-of-plane polarized component.
\end{abstract}
\maketitle

Quantum dots~\cite{arek-book} (QDs) main spectral properties are governed by
size, shape and average chemical composition.~\cite{michler,bimberg-book} 
However the detailed, fine structure of their optical spectra,~\cite{bayer-eh} 
that plays an essential role for applications in quantum optics~\cite{Michler2282,PhysRevLett.86.1502,Stevenson2006}
and information~\cite{Knill2001,Waks2002,RevModPhys.79.135}, is determined by atomic scale details related to microscopic symmetry of underlying lattice,~\cite{singh-bester-eh,zielinski-tallqd,nika-cpqd,our-cpqd} presence of facets,~\cite{Zielinski.PRB.2015} and alloying randomness.~\cite{singh-bester-ordering,singh-bester-lower,Zielinski-natural,Zielinski-elong} 
Regarding potential applications, the bright exciton (BE) recombination in QDs is considered as a tool for generation of entangled photons through biexciton-exciton cascade,~\cite{PhysRevLett.84.2513,prilmuller2018hyperentanglement} whereas the dark exciton (DE) gained attention as a candidate for long-lived, though optically addressable quantum bit.~\cite{PhysRevB.92.201201,PhysRevX.5.011009,Zielinski.PRB.2015,mcfarlane2009gigahertz,xu2008coherent,ohta2018dynamic} The DE is also considered as an auxiliary state for the time-bin entanglement generation scheme.~\cite{PhysRevLett.94.030502,weihs2017time,michler2017quantum}

In self-assembled quantum dots (SADs) efficiency of entanglement generation is limited by distortions~\cite{bayer-eh,singh-bester-lower} of quantum dot (QD) confining potential from an idealized cylindrical symmetry.~\cite{bester-zunger-nair,karlsson,singh-bester-eh,Dupertuis.PRL.2011} The overall low symmetry~\cite{bester-zunger-nair,karlsson,singh-bester-eh,Dupertuis.PRL.2011} induces splitting of optically active excitonic lines, the bright exciton splitting (BES), also known as the fine structure splitting,~\cite{bayer-eh} prohibiting entanglement generation. Since the control of SADs shape is restricted by the character of epitaxial growth, several post-growth methods~\cite{young2005inversion,langbein2004control,akopian2006entangled} has been developed aiming at reduction of the BES, and in particular by utilization of external fields.~\cite{kowalik2005influence,bennett2010electric,Stevenson2006,stevenson2006magnetic,gerardot2007manipulating,seidl2006effect,plumhof2011strain,ding2010tuning,wang2012eliminating,trotta2012universal,trotta2014highly}
Further research focused also on the growth itself, by using ripening~\cite{kors2018telecom} process, droplet epitaxy for low-strain QDs~\cite{abbarchi2010fine} or vapor-liquid-solid (VLS) growth of nanowire QDs (NWQDs).~\cite{bjork,borgstrom-nwqd,dalacu-selective,DalacuUltraClean,maaike-strain,yanase2017single}
The general idea is to restore high symmetry of a QD to reduce its fine structure splitting. This is particularly based on theoretical predictions\cite{singh-bester-eh,karlsson,Dupertuis.PRL.2011} indicating that the triangular ($C_{3v}$) symmetry of a nanostructure will lead to the vanishing BES. Moreover in case of the VLS grown NWQDs the empirical pseudopotential method (EPM) predicts vanishing BES for both pure ($C_{3v}$) and alloyed ($C_1$) NWQDs.~\cite{singh-bester-eh} 
However measurements~\cite{Versteegh2014,huber2014polarization} of the fine structure splitting in alloyed NWQDs show a clear disagreement between experiment and EPM theoretical results.
Notably the EPM predicts~\cite{singh-bester-eh} nearly vanishing fine structure ($0.2\mu$eV) even in heavily alloyed InAs$_{0.25}$P$_{0.75}$ NWQDs, whereas the experiment~\cite{Versteegh2014,huber2014polarization} reveals BES for alloyed NWQDs varying in a broad range of values and reaching up to $16-18$~$\mu$eV.
In case of VLS lithography and NWQDs alloying is unavoidable and origins from the presence of the eutectic growth seed.~\cite{tartakovskii_2012} It leads to a pronounced, up to $80\%$,~\cite{van2009selective,maaike-strain} intermixing of the barrier (InP) material into (InAs) QD region effectively producing heavily alloyed (e.g. InAs$_{0.2}$P$_{0.8}$) NWQDs. In order to achieve the entanglement of the emitted photon pairs in alloyed NWQD researchers thus must do a post-growth~\cite{Versteegh2014} search of low BES samples, somewhat similar to SADs~\cite{akopian2006entangled} and in clear contradiction to the EPM results.

In this work I show by atomistic, empirical tight-binding calculation the fundamental role of composition disorder in NWQDs. In particular pronounced, reaching over $15~\mu$eV BES stemming entirely from the alloy randomness, with no QD shape elongation,~\cite{takagahara,kadantsev-eh,Zielinski-elong} or compositional inhomogeneity.~\cite{huber2014polarization}  
My results are in a very good agreement with the experiment, yet contradicting EPM predictions. Further I propose a scheme of an efficient BES reduction in NWQD via externally applied vertical electric field. Compared to SAD, NWQDs show a very different, Gaussian-like dependence without a lower bound.~\cite{bennett2010electric,singh-bester-lower,plumhof2011strain}
Contrarily the DE splitting in alloyed NWQDs is practically vanishing (below $0.3~\mu$eV) despite mixed chemical composition and alloying. 
The DE oscillator strengths are significantly increased by alloying and changes in QD height, reaching a notable $1/3500$ fraction of the BE
without any NWQD shape alteration.~\cite{Korkusinski.PRB.2013,Zielinski.PRB.2015}
The DE polarization properties are also strongly affected by alloying.
In case of non- or weakly alloyed systems the DE emission is in-plane ("x/y") polarized, whereas in case of strongly alloyed NWQDs this emission gains strong, even dominant, out-of-plane ("z") component.

\begin{figure}
  \begin{center}
  \includegraphics[width=0.5\textwidth]{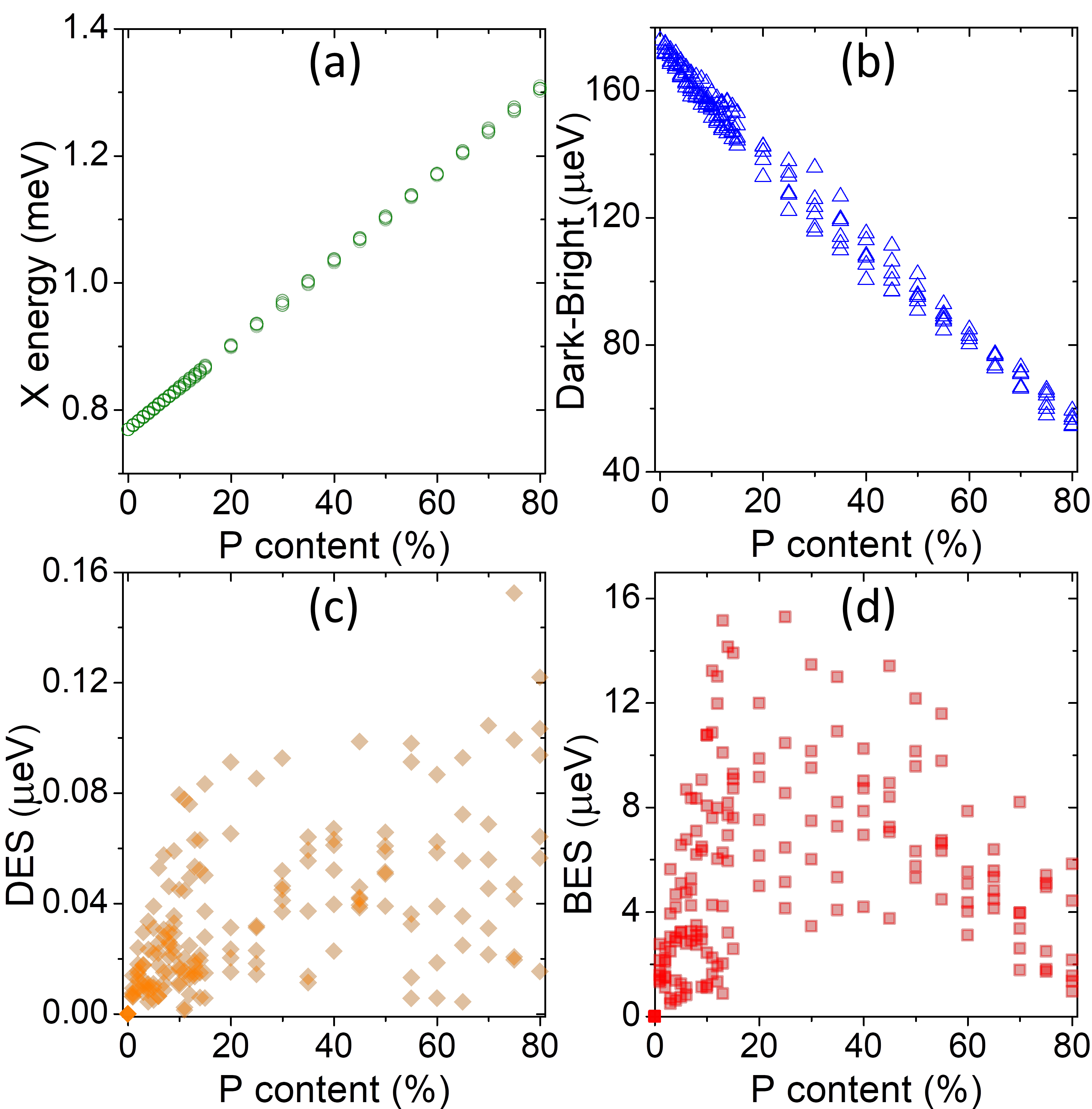}
  \end{center}
  \caption{Exciton ground state energy (a), dark-bright exciton splitting (b), dark (c) and bright (d) exciton splitting as function of phosphorous (P) content in the alloyed NWQD.}
  \label{Pcontent}
\end{figure}

The calculations are performed in a series of computational steps beginning with valence force field~\cite{keating,pryor-zunger,saito-arakawa} approach for strain, empirical tight-binding~\cite{slater-koster,jancu,jaskolski-zielinski-prb06,zielinski-including,zielinski-vbo} (ETB) for the single particle spectra, and configuration interaction (CI) for many-body excitonic properties.~\cite{sheng-cheng-prb2005,zielinski-prb09,rozanski-zielinski,swiderski-zielinski}
Figure~\ref{Pcontent} shows results obtained for an alloyed NWQD as a function of phosphorus ($P$) content. The QD is disk-shaped with $30$~nm diameter and $4.2$~nm height and is embedded in $[111]$ oriented InP host zinc-blend nanowire with diameter of $72$~nm. To account for alloying a uniform composition profile is used mimicking migration of $P$ anions into the QD during the VLS growth.
The overall nanostructure's symmetry is $C_{3v}$ for pure InAs systems and is reduced by alloying to $C_1$. For each $P$ content there are $6$ randomly generated samples corresponding to the same average composition.~\cite{Zielinski-natural}
$P$ content is varied from $10$ to $80\%$ with a $5\%$ step, and from $0\%$ to $10\%$ with a $1\%$ step for greater accuracy.
There are thus $29$ various average compositions times $6$ random samples per composition,  total of $174$ different nanostructures presenting a challenging computational problem.

On Fig.~\ref{Pcontent} (a) the ground excitonic energy increases with $P$ content, from $768$~meV for pure InAs to about $1305$~meV for InAs$_{0.2}$P$_{0.80}$ QD. The growth of the excitonic energy originates from the introduction of the higher band gap energy barrier material into the QD, and to a very good approximation is linear with a $6.7$~meV$/\%$ slope. The spread of excitonic energies due to alloy randomness reaches at most $8$~meV, and is practically non-apparent on Fig.~\ref{Pcontent} (a). On the other hand the dark-bright splitting [Fig.~\ref{Pcontent} (b)], i.e. the energetic difference between the lowest bright and the higher energy dark exciton is reduced with the increasing $P$ content since $P$ add-mixture effectively decreases depth of confinement, and thus so-called isotropic\cite{bayer-eh} electron-hole exchange,  controlling the dark-bright splitting. The spread of calculated values is notable and has maximum of about $20$~meV for $P$ content equal to $30\%$, then it drops with $P$. Intuitively this could be understood in terms of a number of different possible phosphorous atomic arrangements (combinations), growing with $P$, reaching maximum for $P=50\%$ and then reducing again. Speculatively this effect combined with the overall decrease of dark-bright splitting due to $P$ content give maximal spread for $P\approx 30\%$ rather then $P=50\%$.

The impact of alloy randomness on excitonic fine structure [Fig.~\ref{Pcontent} (c) and (d)] is fundamentally stronger. Both the DES and the BES are exactly zero by symmetry for pure InAs NWQDs, yet already a few percent $P$ add-mixture introduces non-negligible splittings.
For the BES [Fig.~\ref{Pcontent} (d)], these splittings reach maximal values of about $15$~$\mu$eV for $P$ content between $10\%$ and $30\%$ and curiously these maximums are quenched again with further alloying, speculatively due to the same mechanism as for the dark-bright splitting. 
For $P$ content in between $20\%$ and $60\%$ minimal values do not drop below about $3$~$\mu$eV, only for lowest $P$ concentration, and for $P$ content over $70\%$ the BES can drop below $1$~$\mu$eV. Not only these values are typically much larger than EPM's predictions,~\cite{singh-bester-eh} but also show pronounced dot-to-dot fluctuations as seen in the experiment.~\cite{Versteegh2014}
Similarly the DES ([Fig.~\ref{Pcontent} (c)] is also triggered by alloying, however this splitting is very small, often below $0.11$~$\mu$eV. Maximal values of the DES increase with $P$ up to about $20\%$, then the trend saturates with only some spikes (of about $0.15$~$\mu$eV) for largest considered $P$ of $75-80\%$. 
The DES distribution is more uniform with no apparent lower bound. 

\begin{figure}
  \begin{center}
  \includegraphics[width=0.5\textwidth]{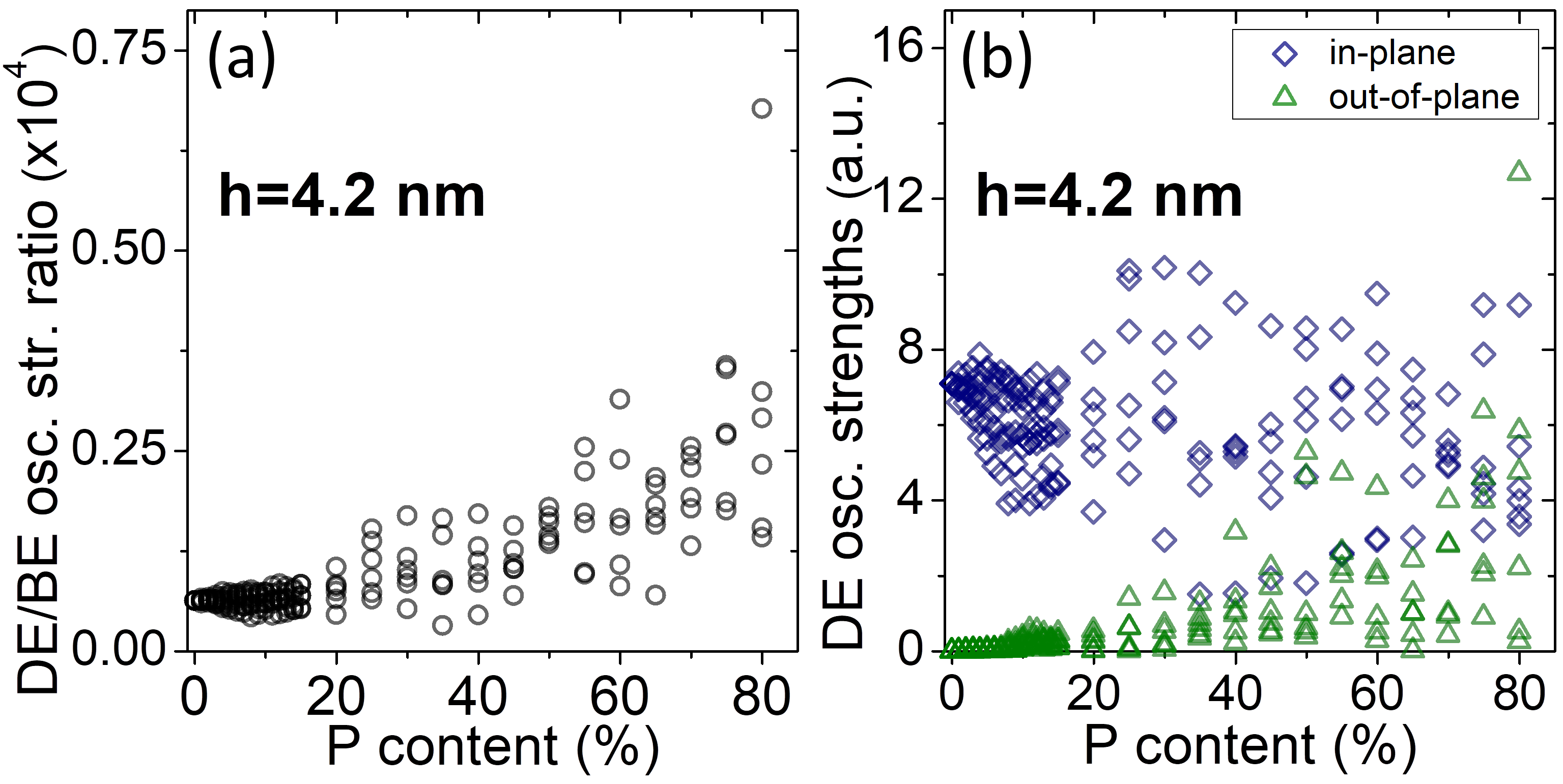}
  \end{center}
  \caption{Ratio of dark to bright exciton oscillator strengths (a) and the dark exciton polarization properties (b) as function of phosphorous (P) content in the alloyed NWQDs. Diamonds(blue)/triangles(green) denote in-plane and out-of-plane polarizations correspondingly.}
  \label{deoscP}
\end{figure}

Alloying has a strong impact on oscillator strengths of excitons, specially on the DE which gains which a substantial optical activity as shown on Fig.~\ref{deoscP} (a), where the ratio of DE to BE oscillator strengths is used as a measure of relative DE optical activity.
For non-alloyed QD the DE oscillator strengths are about $160,000$ 
weaker than the BE. However for alloyed NWQDs this activity is increased by nearly an order of magnitude, up to about $1/15,000$ fraction of the BE, with further significant increase for taller quantum dots, as discussed later in the text. 
Alloying also affects NWQDs polarization properties. 
For the BE, the emitted light is to a very good approximation in QD plane polarized, yet with linear polarization directions randomized from dot-to-dot. The BE emission is thus in-plane polarized with a much weaker ($6$ orders of magnitude) out-of-plane (growth direction, "z") component. For the disk-shaped NWQD there is only little BE's polarization anisotropy, growing with alloying, yet reaching at most about $1\%$ for $P=0.8$. 
The DE exciton spectra of alloyed systems is more curious [Fig.~\ref{deoscP} (b)]. 
For the pure InAs NWQD the DE is still fully in-plane polarized, having exactly the same polarization properties as the BE.~\cite{Dupertuis.PRL.2011} However with the increasing $P$ content the DE gains large out-of-plane component of the emission. 
For the highest considered $P$ content, the out-of-plane oscillator strengths are comparable and can even exceed the in-plane ones. 
This is thus very different from low-symmetry SADs~\cite{Zielinski.PRB.2015} or non-alloyed $C_{3v}$ QDs.~\cite{singh-bester-eh,Dupertuis.PRL.2011} 

\begin{figure}
  \begin{center}
  \includegraphics[width=0.5\textwidth]{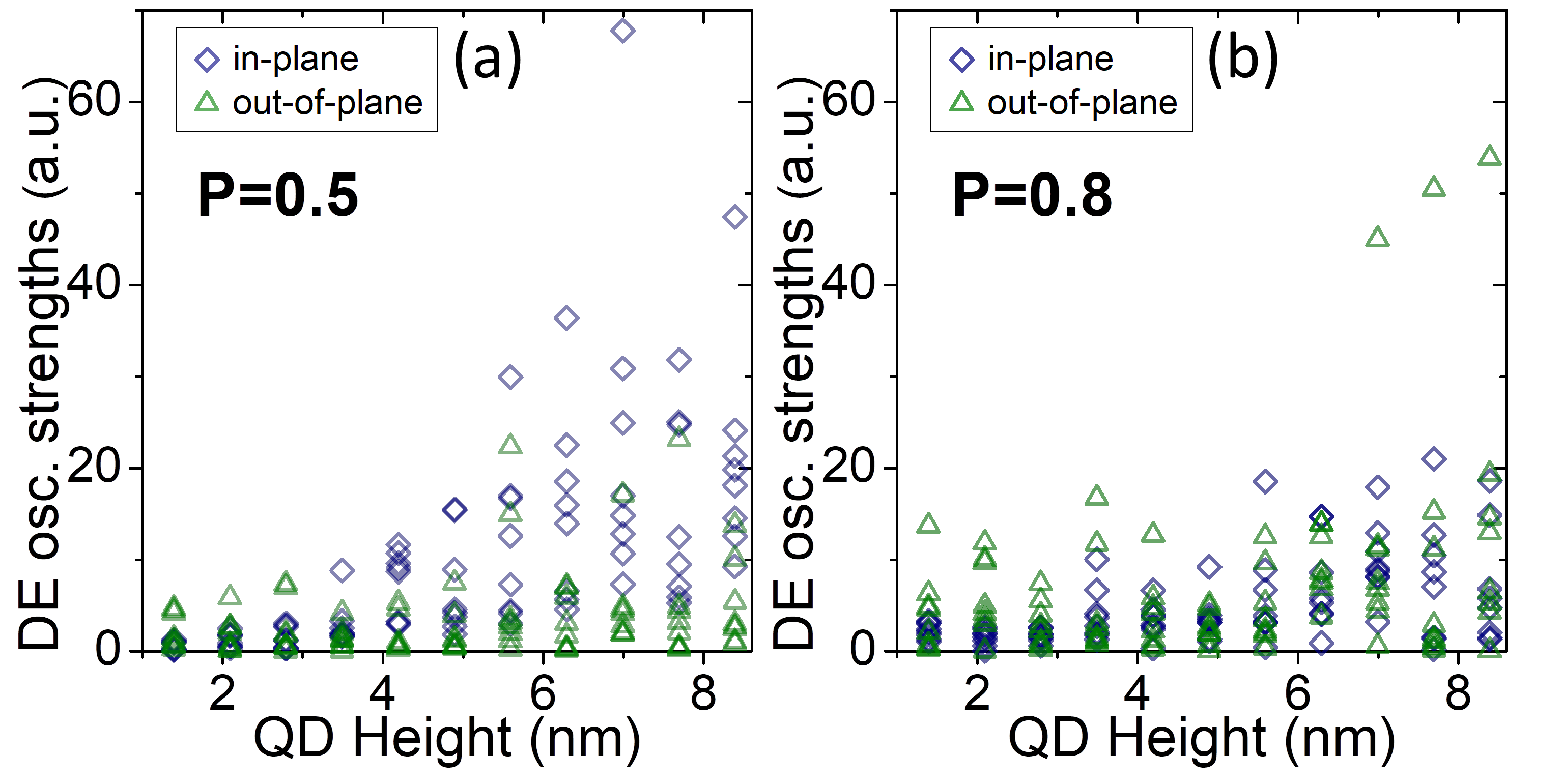}
  \end{center}
  \caption{The dark exciton polarization properties as function of height for alloyed NWQD and two average compositions (a) InAs$_{0.5}$P$_{0.5}$ and (b) InAs$_{0.2}$P$_{0.8}$. Diamonds(blue)/triangles(green) denote in-plane and out-of-plane polarizations correspondingly.}
  \label{deoscH}
\end{figure}

\begin{figure}
  \begin{center}
  \includegraphics[width=0.5\textwidth]{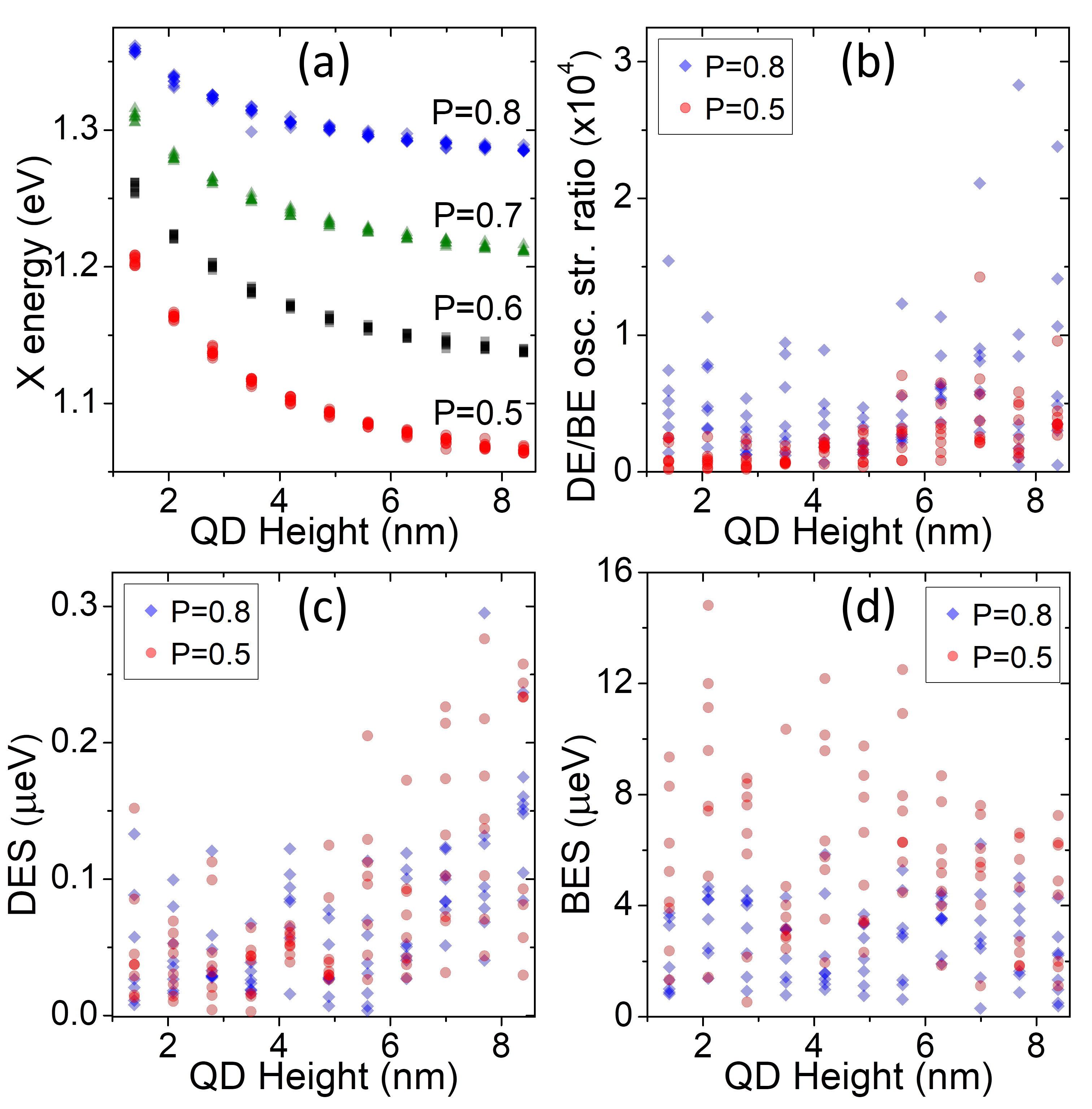}
  \end{center}
  \caption{Exciton ground state energy (a), bright (b) and dark (c) exciton splitting, and dark to bright exciton oscillator strengths ratio (d) as function of alloyed NWQD height; different symbols/colors denote phosphorous contents.}
  \label{H}
\end{figure}

It is also curious to check how spectra of excitons in alloyed NWQDs depend on the growth direction confinement. This is shown on Fig.~\ref{deoscH} and Fig.~\ref{H} where the height of NWQDs varies from $1.4$~nm to $8.4$~nm ($11$ different cases). All other QD and nanowire dimensions are the same as above. Four different $P$ concentrations were considered ($P$=$50\%$ and $80\%$, as well as $60\%$, $70\%$ for comparison). There were $8$ random samples generated for each average composition, leading to a formidable problem of ($11\times 4 \times 8$) $352$ separate atomistic computations.

For InAs$_{0.5}$P$_{0.5}$ NWQDs and heights lower than $3$~nm, the DE is mostly out-of-plane polarized with a weak in-plane components [Fig.~\ref{deoscH} (a)]. This is quite curious and it seems that due to randomness the DE polarization properties in flat alloyed NWQDs resemble more $C_{2v}$ SADs spectra,~\cite{Dupertuis.PRL.2011} with DE "z" polarized emission, rather than $C_{3v}$ pure InAs NWQDs with "x/y" polarization. However with the increase of QD height, the in-plane component grows faster then the out-of-plane one, and the in-plane polarization dominates the tall QD DE spectra of the $P=0.5$ case. For heavily alloyed $P=0.8$ systems and small QD heights, the DE emission is mostly out-of-plane polarized [Fig.~\ref{deoscH} (a)]. For taller QDs and high $P=0.8$ content both components are comparable. However there are several notable cases where the out-of-plane polarization becomes dominant for $h>7$~nm, again very distinct from non-alloyed $C_{3v}$ systems.

Further studies of QDs height dependence are shown on Figure~\ref{H}. Here for completeness Figure~\ref{H} (a) presents the ground excitonic state energy evolution as a function of both height and $P$ content. The increased QD height reduces confinement and decreases excitonic energy, whereas the $P$ content shifts up excitonic energy and flattens the trends. The spread of calculated values due to alloy randomness is again relatively small ($\approx10$~meV). 
Figure~\ref{H} (b) shows the DE activity with respect to BE, similar to Fig.~\ref{deoscP} (a), yet this time as a function of QD height.
For $P=0.5$ there is about five-fold increase of the DE optical activity as a result of reduced confinement. Similarly for $P=0.8$ the DE/BE oscillator strengths ratio is much larger for taller QDs than for flat ones, with an exception of really flat, several monolayer ($<2$ nm) thick QDs, where DE/BE ratio is also increasing. Notably for the highly alloyed and high aspect ratio, tall QDs the DE can reach significant optical activity being $1/3500$ fraction of the BE.
Further, Figure~\ref{H} (c) and (d) present DES and BES correspondingly as a function of QD height. The DES spread increases with QD height, reaching about $0.15$~$\mu$eV for $h<5$~nm and about $0.3$~$\mu$eV for taller systems. 
The lower bound of the DE spin's coherent precession time~\cite{Poem.Nature.2010}  (Planck’s constant divided by eigenstates’ energy difference) will thus typically exceed $14$~ns, an important figure for potential DE applications.~\cite{PhysRevB.92.201201} For the DES there is no also apparent difference between $P$ cases, coherent with earlier discussions. 
Contrarily the BES shows nearly no dependence on the QD height, and a strong dependence on $P$ content. For $P=0.5$, the BES is generally $2$ to $3$ times larger than for $P=0.8$. Thus the BES is reduced with a large alloying, consistent with earlier results. Notably the BES varies greatly between distinct QDs and only small group of QDs will have the BES below the conventional threshold of $1$~$\mu$eV, consistent with experimental findings and contrasting the EPM predictions.

\begin{figure}
  \begin{center}
  \includegraphics[width=0.5\textwidth]{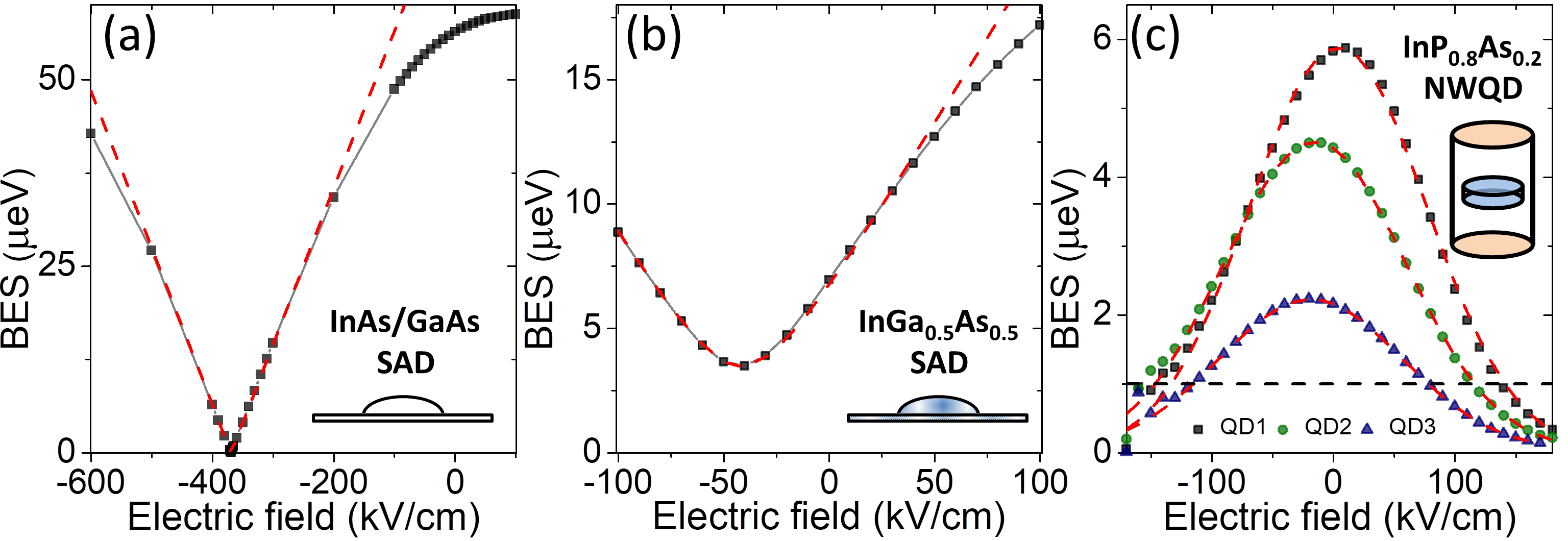}
  \end{center}
  \caption{The bright exciton splitting as a function of vertical electric field for (a) non-alloyed InAs/GaAs SAD, (b) alloyed InGa$_{0.5}$As$_{0.5}$/GaAs SAD, (c) for three different alloyed NWQDs of the same average composition InP$_{0.8}$As$_{0.2}$/InP.}
  \label{besall}
\end{figure}

Control of the BES with external field is important for applications.~\cite{prilmuller2018hyperentanglement} Figure~\ref{besall} shows the effect of a vertical electric field~\cite{marcet2010vertical} applied to NWQDs as well as SADs for comparison.~\cite{bennett2010electric}
Two lens-shaped SADs are considered: a non-alloyed InAs of $C_{2v}$ symmetry and alloyed InGa$_{0.5}$As$_{0.5}$ of $C_1$ symmetry. SADs dimensions are identical with diameters of $25$~nm and heights equal to $3.5$~nm. Both SADs are embedded in GaAs barrier and are placed on 2 monolayer thick ($0.6$~nm) InAs wetting layer.
As for the NWQDs, three different InP$_{0.8}$As$_{0.2}$ alloyed dots are shown on Figure~\ref{besall} (c). These NWQDs have identical dimensions (h=$4.2$~nm, d=$30$~nm) and average composition, the only difference between them being random alignment of $P$ atoms. Separate ETB and CI calculations were performed for total of $215$ different cases.

For the $C_{2v}$ SAD the zero field BES [Figure~\ref{besall} (a)] is large and equal to $56.4$~$\mu$eV. 
Field affect this splitting and actually even reverse the order of BE lines.~\cite{bennett2010electric} Alloying reduces~\cite{langbein2004control} BES in SADs, and for InGa$_{0.5}$As$_{0.5}$ the BES at zero field is equal to about $7$~$\mu$eV  [Figure~\ref{besall} (b)], whereas the field can reduce the BES to its lower bound~\cite{bennett2010electric,singh-bester-eh} of $3.5$~$\mu$eV. These results corresponds well to the experiment for similar QDs.~\cite{bennett2010electric} Atomistic calculation can also be very well fit~\footnote{Results of fitting using Ref.~\cite{bennett2010electric} notation: for $C_{2v}$ InAs/GaAs QD $s_0=0$ and $\gamma=0.21$~$\mu$eV kV$^{-1}/$cm; for alloyed InGa$_{0.5}$As$_{0.5}$ $s_0=3.49$~$\mu$eV, and $\gamma=0.14$~$\mu$eV kV$^{-1}/$cm.} [red/dashed lines on Fig.~\ref{besall} (a) and (b)] to a phenomenological model proposed in Ref.~\cite{bennett2010electric}.

In SADs electric field tuning of the BES stems from a dipole moment along the growth direction~\cite{barker2000theoretical,bennett2010electric} coming from asymmetric shape of SADs.~\cite{barker2000theoretical}  In NWQDs there is no shape asymmetry. There is an inversion asymmetry~\cite{ZielinskiSubstrate2012,zielinski-tallqd} related to the $[111$] growth, that spatially shifts  the electron and the hole.~\cite{swiderski-zielinski}. This effects however does not trigger the BES, which is exactly zero for $C_{3v}$, and the BES is only due to alloying. Moreover NWQDs show a peculiar field BES evolution  [Fig.~\ref{besall} (c)]. A particle in the box model, assuming infinite square well, relating the BES to the oscillator strength,~\cite{takagahara} and using perturbation theory up to the $2$-order, gives the BES dependence as: $s_0-\alpha F^2 + \beta F^4$, where $s_0$ is the zero field $F$ splitting, $\alpha$ and $\beta$ are constants. For small fields this matches well the trend shown on Fig.~\ref{besall} (c). 
Going beyond this crude approach and assuming harmonic-oscillator type confinement~\cite{arek-book,kadantsev-eh} one gets Gaussian-like field BES dependence. 
This shape allows for very good fits to atomistic calculations [red/dashed lines on Fig.~\ref{besall} (c)] with the BES not centered at zero field, again due to alloying.
For small fields the Gaussian is consistent with a simple model, the latter resembling the first terms of the Gaussian series expansion.
QDs with smaller zero field BES reach threshold of $1$~$\mu$eV [black/dashed line on Fig.~\ref{besall} (c)] at smaller applied fields. The BES field reduction is associated with an increasing electron-hole spatial separation and thus the decreasing BE oscillator strengths, with the oscillator strengths evolution in field resembling very much that of the BES. For NWQDs considered on Fig.~\ref{besall} (c) there is about a four-fold reduction of the oscillator strength for BES reaching $1$~$\mu$eV threshold, thus NWQDs with field reduced BES remain to be optically active.

In conclusion, it has been shown shown that alloying in nanowire quantum dots is responsible for a non-vanishing bright exciton splitting occurring merely due to alloy randomness without any shape deformation, off-center quantum dot position, nor non-uniform composition. The BES depends highly on the composition intermixing and varies considerably between individual dots. The splitting can be controlled by vertical electric field, with a field dependence very different from SADs. The dark exciton properties in nanowire quantum dots are sensitive to the alloying as well. The DE gains notable optical activity that grows with alloying and quantum dot height. Heavily alloyed nanowire quantum dots can have strongly out-of-plane polarized dark exciton emission, whereas in weakly alloyed NWQDs the DE will be in-plane polarized. 

The author would like to thank prof. David Gershoni for discussions prior to this work.
The support from the Polish National Science Centre based on decision No. 2015/18/E/ST3/00583 is kindly acknowledged.

\bibliography{main}

\end{document}